\documentclass[11pt]{article}

\usepackage{acl}

\usepackage{times}
\usepackage{latexsym}
\usepackage[T1]{fontenc}
\usepackage[utf8]{inputenc}
\usepackage{microtype}
\usepackage{inconsolata}

\usepackage{graphicx}
\usepackage{float}
\usepackage{booktabs}
\usepackage{amsmath}
\usepackage{amssymb}
\usepackage{multirow}
\usepackage{xcolor}
\usepackage{colortbl}
\usepackage{tikz}
\usepackage{pgfplots}
\pgfplotsset{compat=1.17}
\usetikzlibrary{shapes.geometric, arrows, positioning, calc, patterns, shadows, decorations.pathreplacing}

\definecolor{humancolor}{RGB}{65, 105, 225}
\definecolor{machinecolor}{RGB}{220, 20, 60}
\definecolor{openaicolor}{RGB}{116, 185, 255}
\definecolor{llamacolor}{RGB}{255, 177, 66}
\definecolor{qwencolor}{RGB}{85, 239, 196}

\title{UCSC-NLP at SemEval-2026 Task 13: Multi-View Generalization and Diagnostic Analysis of Machine-Generated Code Detection}

\author{
Kargi Chauhan\thanks{Both authors contributed equally.} \\
University of California, Santa Cruz \\
\texttt{kchauha3@ucsc.edu}
\And
Sadiba Nusrat Nur\footnotemark[1] \\
University of California, Santa Cruz \\
\texttt{sanur@ucsc.edu}
}

\begin{document}
\maketitle

\begin{abstract}
With the rapid growth of large language models for code generation, distinguishing between human-written and AI-generated code has become increasingly critical for academic integrity, hiring evaluations, and software security. We present our system for SemEval-2026 Task 13: Multilingual Machine-Generated Code Detection, participating in Subtask~A (binary detection) and Subtask~B (multi-class attribution across 10 LLM families). For Subtask A, we fine-tune UniXcoder-base with a multi-view training framework that promotes generator-invariant representations. The framework combines domain-specific structural prefixes, delexicalization with symmetric KL consistency loss, token dropout, and mixed-content augmentation. Our system achieves 0.993 macro F1 on validation and 0.845 macro F1 on the test set, which spans unseen languages and domains. For Subtask~B, we show that severe class imbalance (88.4\% human code, 221:1 majority-to-minority ratio) causes catastrophic minority-class failure under standard fine-tuning, with macro F1 collapsing to 0.086 despite 88.4\% accuracy. A class-weighted extension trained for 3 epochs recovers macro F1 to 0.345 (+301\% relative), confirming that multi-class attribution requires imbalance-aware training strategies.\footnote{Code: https://github.com/Kargichauhan/Detecting-Machine-Generated-Code}
\end{abstract}

\section{Introduction}

Large language models (LLMs) have reshaped software development. AI-assisted tools such as GitHub Copilot, used by over 1.8 million developers \citep{dohmke2023}, are now embedded in educational and professional workflows. While improving productivity, they raise concerns about authorship, academic integrity, intellectual property, and security auditing \citep{pearce2022asleep}. Consequently, distinguishing human-written from machine-generated code has become both critical and technically challenging, as detectors often overfit to generator-specific artifacts and degrade under prompt variation or minor edits \citep{pan2024assessing, liu2024detecting}.
SemEval-2026 Task 13 \citep{orel-etal-2026-semeval-task13} builds upon 
the DROID resource suite \citep{orel-etal-2025-droid}, a multilingual, multi-generator benchmark designed to evaluate robustness under distribution shift. Prior work shows that detectors trained on limited languages or domains generalize poorly to unseen settings \citep{orel2025codet, pan2024assessing}. These issues are exacerbated in fine-grained attribution, where severe class imbalance leads to systematic neglect of minority generator classes.
We frame this task as a domain generalization problem under lexical, structural, and stylistic shifts. To promote generator-invariant representations, we adopt: (i) domain-specific structural prefixes for content-type alignment \citep{ganin2016domain}, (ii) delexicalization with symmetric KL consistency regularization \citep{miyato2018vat}, (iii) token dropout to reduce reliance on spurious cues \citep{wei2019eda}, and (iv) mixed-content augmentation for distributional robustness \citep{arjovsky2019irm}. We further show that multi-class attribution collapses under severe imbalance, where high accuracy masks near-complete minority-class failure. Our work makes two distinct contributions: (1) a multi-view generalization framework for robust binary detection under distribution shift (Subtask A), 
and (2) a diagnostic analysis demonstrating that standard fine-tuning catastrophically fails under extreme class imbalance in multi-class attribution (Subtask B).

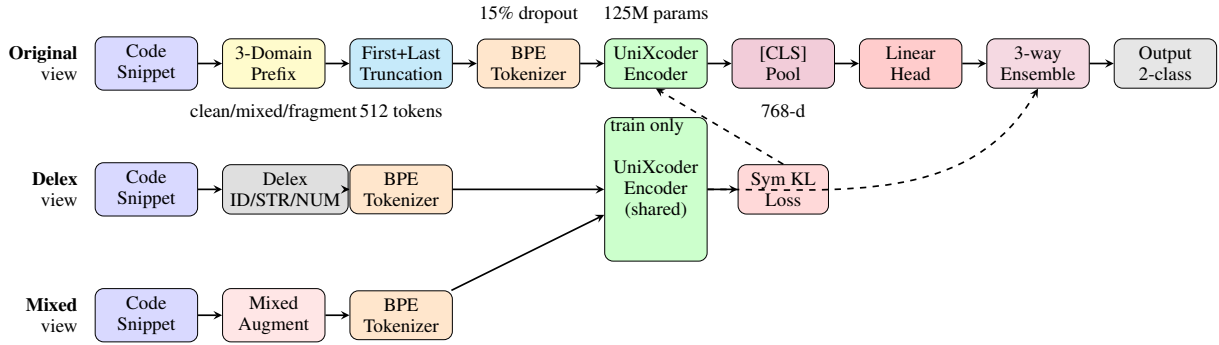
\begin{figure*}[h]
\centering
\resizebox{\textwidth}{!}{%
\begin{tikzpicture}[
    node distance=0.4cm,
    box/.style={rectangle, draw, rounded corners, minimum width=1.7cm, minimum height=0.7cm, align=center, font=\small},
    smallbox/.style={rectangle, draw, rounded corners, minimum width=1.5cm, minimum height=0.7cm, align=center, font=\small},
    arrow/.style={->, thick, >=stealth},
    dasharrow/.style={->, thick, >=stealth, dashed}
]

\node[box, fill=blue!15] (input) {Code\\Snippet};
\node[box, fill=yellow!25, right=0.4cm of input] (prefix) {3-Domain\\Prefix};
\node[box, fill=cyan!20, right=0.4cm of prefix] (firstlast) {First+Last\\Truncation};
\node[box, fill=orange!20, right=0.4cm of firstlast] (token) {BPE\\Tokenizer};
\node[box, fill=green!20, right=0.4cm of token] (bert) {UniXcoder\\Encoder};
\node[box, fill=purple!20, right=0.4cm of bert] (cls) {[CLS]\\Pool};
\node[box, fill=red!20, right=0.4cm of cls] (head) {Linear\\Head};
\node[box, fill=purple!15, right=0.4cm of head] (ensemble) {3-way\\Ensemble};
\node[box, fill=gray!20, right=0.4cm of ensemble] (out) {Output\\2-class};

\node[box, fill=blue!15, below=1.2cm of input] (input2) {Code\\Snippet};
\node[box, fill=gray!25, right=0.4cm of input2] (delex) {Delex\\ID/STR/NUM};
\node[box, fill=orange!20] (token2) at (firstlast |- input2) {BPE\\Tokenizer};

\node[box, fill=blue!15, below=1.2cm of input2] (input3) {Code\\Snippet};
\node[box, fill=pink!40, right=0.4cm of input3] (mix) {Mixed\\Augment};
\node[box, fill=orange!20] (token3) at (firstlast |- input3) {BPE\\Tokenizer};

\node[box, fill=green!20, minimum height=2.4cm] (bert2) at (bert |- input2) {UniXcoder\\Encoder\\(shared)};

\node[smallbox, fill=red!15] (kl) at (cls |- input2) {Sym KL\\Loss};

\draw[arrow] (input) -- (prefix);
\draw[arrow] (prefix) -- (firstlast);
\draw[arrow] (firstlast) -- (token);
\draw[arrow] (token) -- (bert);
\draw[arrow] (bert) -- (cls);
\draw[arrow] (cls) -- (head);
\draw[arrow] (head) -- (ensemble);
\draw[arrow] (ensemble) -- (out);

\draw[arrow] (input2) -- (delex);
\draw[arrow] (delex) -- (token2);
\draw[arrow] (token2) -- (bert2);
\draw[arrow] (bert2) -- (kl);

\draw[arrow] (input3) -- (mix);
\draw[arrow] (mix) -- (token3);
\draw[arrow] (token3) -- (bert2);

\draw[dasharrow] (kl.north) -- (bert.south);
\node[font=\footnotesize] at ($(kl.north)!0.5!(bert.south) + (-1.2cm, 0)$) {train only};

\draw[dasharrow] (bert2.east) to[out=0, in=240] (ensemble.south);

\node[below=0.12cm of prefix, font=\footnotesize] {clean/mixed/fragment};
\node[below=0.12cm of firstlast, font=\footnotesize] {512 tokens};
\node[above=0.12cm of bert, font=\footnotesize] {125M params};
\node[below=0.12cm of cls, font=\footnotesize] {768-d};
\node[above=0.12cm of token, font=\footnotesize] {15\% dropout};

\node[left=0.15cm of input,  font=\footnotesize, align=right] {\textbf{Original}\\view};
\node[left=0.15cm of input2, font=\footnotesize, align=right] {\textbf{Delex}\\view};
\node[left=0.15cm of input3, font=\footnotesize, align=right] {\textbf{Mixed}\\view};

\end{tikzpicture}%
}
\caption{Subtask~A architecture. Three views of each input are processed through a shared UniXcoder encoder; symmetric KL loss enforces prediction consistency across views during training, while inference averages logits as a zero-cost ensemble.}
\label{fig:architecture_a}
\end{figure*}

\begin{figure*}[t]
\centering
\resizebox{0.9\textwidth}{!}{%
\begin{tikzpicture}[
    node distance=0.6cm,
    box/.style={rectangle, draw, rounded corners, minimum width=2cm, minimum height=0.8cm, align=center, font=\small},
    arrow/.style={->, thick, >=stealth}
]
\node[box, fill=blue!15] (input) {Code Snippet};
\node[box, fill=orange!20, right=of input] (token) {BPE Tokenizer};
\node[box, fill=green!20, right=of token] (bert) {CodeBERT Encoder};
\node[box, fill=purple!20, right=of bert] (pool) {[CLS] Pooling};
\node[box, fill=red!20, right=of pool] (class) {Linear Classifier};
\node[box, fill=gray!20, right=of class] (out) {11-Class Output};

\draw[arrow] (input) -- (token);
\draw[arrow] (token) -- (bert);
\draw[arrow] (bert) -- (pool);
\draw[arrow] (pool) -- (class);
\draw[arrow] (class) -- (out);

\node[below=0.2cm of token, font=\scriptsize] {$L = 512$};
\node[below=0.2cm of bert, font=\scriptsize] {CodeBERT-base (125M)};
\node[below=0.2cm of pool, font=\scriptsize] {$\mathbf{h}_{\texttt{[CLS]}} \in \mathbb{R}^{768}$};
\node[below=0.2cm of class, font=\scriptsize] {Human + 10 LLM Families};

\end{tikzpicture}%
}
\caption{Subtask~B architecture. A standard CodeBERT encoder with softmax classification head serves as a diagnostic baseline to isolate the effect of class imbalance on 11-way generator attribution.}
\label{fig:architecture_taskb}
\end{figure*}
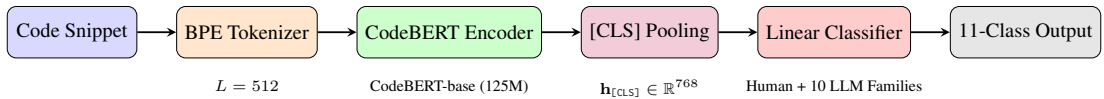

\vspace{-0.6em}
\section{Background and Related Work}

\subsection{Machine-Generated Code Detection}

Early work on machine-generated code detection adapts text-based AI detection methods to programming languages. GPTSniffer \citep{nguyen2024gptsniffer} introduced CodeBERT-based classification and achieved strong performance on ChatGPT-generated code, but showed limited generalization across generators. CodeGPTSensor \citep{xu2025contrastive} mitigated this using contrastive learning to better separate human-written and machine-generated representations.

Prior studies highlight the difficulty of the task. General-purpose AI detectors perform near chance level on code \citep{pan2024assessing}, and learned stylistic cues are often generator-specific and brittle to prompt variation or post-editing \citep{liu2024detecting}. Alternative signals, such as entropy-based detection \citep{mitchell2023detectgpt}, repetition statistics \citep{gehrmann2019gltr}, and likelihood-based methods \citep{kirchenbauer2023watermark}, are largely designed for natural language and degrade on source code due to its syntactic regularity. Recent benchmarks \citep{wang2024ai} therefore emphasize robustness across programming languages and generators as a key challenge.

\subsection{Pretrained Code Models}

CodeBERT \citep{feng-etal-2020-codebert} introduced bimodal pretraining on natural language and code, enabling strong downstream classification. Subsequent models incorporate richer structural signals: GraphCodeBERT \citep{guo-etal-2021-graphcodebert} integrates data-flow graphs, while UniXcoder \citep{guo-etal-2022-unixcoder} unifies encoder-only, decoder-only, and encoder--decoder objectives within a structure-aware pretraining framework, demonstrating strong cross-task transfer.

Sequence-to-sequence approaches such as CodeT5 \citep{wang2021codet5} and PLBART \citep{ahmad2021plbart} further extend pretraining for code generation and understanding. Despite increasing architectural complexity, encoder-based models remain competitive for classification under moderate computational budgets \citep{kanade2020learning}.

\section{Dataset Analysis}

\subsection{Subtask A: Cross-Lingual Generalization}

The binary dataset exhibits near-ideal label balance (47.7\% human, 52.3\% machine) but presents a critical language and domain mismatch between training and evaluation. Training is restricted to Python (91\%), C++, and Java, all from the algorithmic domain. The test set, however, spans unseen languages (JavaScript, C\#, PHP, Go, C) and unseen domains (Research, Production), with 33.4\% of samples being mixed-content snippets of unknown type. This train--test distribution shift, rather than label imbalance, is the central challenge of Subtask~A (Table~\ref{tab:task_a_dist}).

\begin{table}[t]
\caption{Subtask~A language and domain distribution. The test set introduces five unseen languages and two unseen domains absent from training.}
\centering
\small
\setlength{\tabcolsep}{4pt}
\begin{tabular}{lllc}
\toprule
\textbf{Split} & \textbf{Lang.} & \textbf{Domain} & \textbf{\%} \\
\midrule
\multirow{3}{*}{Train}
 & Python & Algo. & 91\% \\
 & C++    & Algo. & 5\%  \\
 & Java   & Algo. & 4\%  \\
\midrule
\multirow{8}{*}{Test}
 & Python      & Algo.      & 28.3\% \\
 & JavaScript  & Mixed      & 13.3\% \\
 & Java        & Mixed      & 12.0\% \\
 & C\#         & Res./Prod. & 5.4\%  \\
 & PHP         & Res./Prod. & 5.4\%  \\
 & C++         & Algo.      & 1.7\%  \\
 & Go          & Res./Prod. & 0.5\%  \\
 & Unknown     & Mixed      & 33.4\% \\
\bottomrule
\end{tabular}
\label{tab:task_a_dist}
\end{table}

\subsection{Subtask B: Severely Imbalanced Multi-Class}

Table~\ref{tab:task_b_dist} reveals the core challenge of Subtask~B: the training data is dominated by human-written code (442K samples, 88.4\%), while machine-generated samples are spread across 10 LLM families with counts ranging from 2K to 10K. The imbalance ratio between the largest class (Human) and the smallest (BigCode, Gemma) reaches \textbf{221:1}, fundamentally biasing gradient-based optimization toward the majority class. Standard cross-entropy training under this distribution causes the model to effectively predict ``human'' for the vast majority of inputs regardless of true label.

\begin{table}[t]
\caption{Subtask~B class distribution. The 221:1 imbalance ratio between the majority class and the smallest minority classes poses a fundamental optimization challenge.}
\centering
\small
\begin{tabular}{lrr}
\toprule
\textbf{Class} & \textbf{Samples} & \textbf{\% of Train} \\
\midrule
Human          & 442,000 & 88.4\% \\
OpenAI         &  10,000 &  2.0\% \\
Meta-LLaMA     &   8,000 &  1.6\% \\
IBM-Granite    &   8,000 &  1.6\% \\
Qwen           &   8,000 &  1.6\% \\
Phi            &   5,000 &  1.0\% \\
DeepSeek-AI    &   4,000 &  0.8\% \\
Mistral        &   4,000 &  0.8\% \\
01-ai          &   3,000 &  0.6\% \\
BigCode        &   2,000 &  0.4\% \\
Gemma          &   2,000 &  0.4\% \\
\midrule
\textbf{Total} & \textbf{500,000} & \textbf{100\%} \\
\bottomrule
\end{tabular}
\label{tab:task_b_dist}
\end{table}

\section{Methodology}

\subsection{Subtask A: Multi-View Generalization Framework}

Our initial CodeBERT baseline achieved near-perfect validation F1 but failed 
to generalize to the test set due to the language and domain mismatch 
described in Section~3.1. We address this by fine-tuning UniXcoder-base 
\citep{guo-etal-2022-unixcoder} with a multi-view training framework 
(Figure~\ref{fig:architecture_a}). We introduce four targeted modifications, 
each addressing a specific failure mode identified in Section~3.1.

 \textbf{3-Domain Structural Prefix.} The test set contains 33.4\% mixed-content snippets entirely absent from training, causing the model to apply incorrect decision boundaries to unseen content types. We address this by prepending a structured prefix classifying each snippet into \texttt{clean}, \texttt{mixed}, or \texttt{fragment} based on the ratio of syntactic to natural-language lines, along with structural features (loops, functions, classes).

 \textbf{Delexicalization with Symmetric KL Consistency Loss.} Lexical surface cues --- variable names, library identifiers, string constants --- vary systematically across programming languages and do not transfer to unseen domains. We replace all identifiers, string literals, and numeric literals with generic tokens (\texttt{ID}, \texttt{STR}, \texttt{NUM}) and enforce prediction consistency between the original and delexicalized views via symmetric KL divergence. 

\textbf{Token Dropout.} Language-specific keywords act as spurious features that the model can exploit on the training distribution but that do not generalize to unseen languages. We randomly replace 15\% of non-special tokens with \texttt{[MASK]} during training, simulating exposure to unknown syntax.

\textbf{Mixed-Content Augmentation.} Training data contains no mixed-content samples, yet 33.4\% of test samples are mixed code--text snippets of unknown type. We randomly inject generic text fragments into training snippets at 40\% probability with a softer consistency loss, directly bridging this distributional gap. The combined training loss is:

\vspace{-1.2em}
\begin{equation}
\mathcal{L} = \mathcal{L}_{\text{CE}} + \lambda \cdot \text{SymKL}(p \| q_{\text{delex}})
+ \frac{\lambda}{2} \cdot \text{SymKL}(p \| q_{\text{mixed}})
\end{equation}
At inference, logits from all three views (original, delexicalized, mixed) are averaged as a zero-cost test-time ensemble.

\subsection{Subtask B: Diagnostic Baseline and Class-Weighted Extension}

Subtask~B is formulated as an 11-class classification problem, where each code snippet is labeled as either human-written or generated by one of 10 LLM families. We fine-tune \texttt{microsoft/codebert-base} \citep{feng-etal-2020-codebert} with a linear softmax classification head over the pooled \texttt{[CLS]} representation ($\mathbf{h}_{\texttt{[CLS]}} \in \mathbb{R}^{768}$), using a maximum sequence length of 512 tokens (Figure~\ref{fig:architecture_taskb}).

We present two systems. First, an \textbf{unweighted diagnostic baseline} that deliberately avoids class reweighting, focal loss, or oversampling to isolate the impact of extreme label skew the divergence between macro F1 and accuracy directly measures minority-class collapse. Second, a \textbf{class-weighted competitive system} that applies inverse-frequency weighting to the cross-entropy loss, upweighting each minority LLM class proportionally to its underrepresentation. Both systems are trained for 3 epochs.

\subsection{Training Configuration}

Table~\ref{tab:hyperparams} summarizes hyperparameters for both subtasks. For Subtask~A, we use first+last truncation (concatenating the first and last 256 tokens to form a 512-token input) to preserve both opening signatures and closing patterns.

\begin{table}[t]
\centering
\small
\begin{tabular}{lcc}
\toprule
\textbf{Hyperparameter} & \textbf{Task A} & \textbf{Task B} \\
\midrule
Backbone              & UniXcoder-base   & CodeBERT-base \\
Max sequence length   & 512 (first+last) & 512 \\
Learning rate         & $2 \times 10^{-5}$ & $2 \times 10^{-5}$ \\
Batch size            & 16  & 32  \\
Epochs                & 1   & 3   \\
Warmup steps          & 470 & 500 \\
Weight decay          & 0.01 & 0.01 \\
Label smoothing       & 0.1  & ---  \\
KL weight $\lambda$   & 0.5  &  0  \\
Token dropout rate    & 15\% &  0  \\
Class weighting       & 0  & Inv.-freq. \\
\bottomrule
\end{tabular}
\caption{Training hyperparameters. Task~A uses a single epoch with multi-view regularization; Task~B trains for 3 epochs, with the competitive system applying inverse-frequency class weighting.}
\label{tab:hyperparams}
\end{table}

\section{Results}
\label{sec:results}

\subsection{Subtask A: Binary Classification}

Table~\ref{tab:task_a_results} presents our Subtask~A results. The system achieves 0.993 macro F1 on the in-distribution validation set and 0.845 on the out-of-distribution test set. The 14.8-point gap reflects the difficulty of cross-lingual and cross-domain generalization, though the test performance remains strong given that training uses only three languages in a single domain. Per-class test performance is balanced: Human (P\,=\,0.920, R\,=\,0.760, F1\,=\,0.833) and Machine (P\,=\,0.794, R\,=\,0.938, F1\,=\,0.860), indicating no systematic bias toward either class.

\begin{table}[t]
\caption{Subtask~A results. The val--test gap reflects cross-lingual and cross-domain distribution shift, not overfitting.}
\centering
\small
\begin{tabular}{lccc}
\toprule
\textbf{Split} & \textbf{Accuracy} & \textbf{Macro F1} & \textbf{AUC} \\
\midrule
Validation & 0.993 & 0.993 & 0.999 \\
Test       & 0.892 & 0.845 & 0.851 \\
\bottomrule
\end{tabular}
\label{tab:task_a_results}
\end{table}

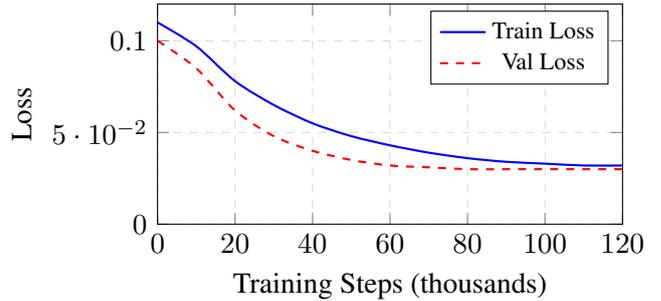
\begin{figure}[t]
\centering
\begin{tikzpicture}
\begin{axis}[
    width=\columnwidth,
    height=4.5cm,
    xlabel={Training Steps (thousands)},
    ylabel={Loss},
    xmin=0, xmax=120,
    ymin=0, ymax=0.12,
    legend pos=north east,
    legend style={font=\small},
    grid=major,
    grid style={dashed, gray!30},
]
\addplot[color=blue, thick, mark=none, smooth] coordinates {
    (0, 0.11) (10, 0.097) (20, 0.078) (30, 0.065)
    (40, 0.055) (50, 0.048) (60, 0.043) (70, 0.039)
    (80, 0.036) (90, 0.034) (100, 0.033) (110, 0.032) (120, 0.032)
};
\addplot[color=red, thick, mark=none, smooth, dashed] coordinates {
    (0, 0.10) (10, 0.085) (20, 0.062) (30, 0.048)
    (40, 0.040) (50, 0.035) (60, 0.032) (70, 0.031)
    (80, 0.030) (90, 0.030) (100, 0.030) (110, 0.030) (120, 0.030)
};
\legend{Train Loss, Val Loss}
\end{axis}
\end{tikzpicture}
\caption{Subtask~A training dynamics. Validation loss closely tracks training loss, indicating stable convergence without overfitting across 120K steps.}
\label{fig:task_a_loss}
\end{figure}

Figure~\ref{fig:task_a_loss} confirms healthy training dynamics: validation loss tracks training loss throughout optimization with no divergence, indicating that the multi-view regularization framework prevents overfitting despite training for only a single epoch.

\subsection{Subtask B: Multi-Class Attribution}
\label{sec:results_b}


\begin{table}[h!]
\caption{Subtask~B results. The diagnostic baseline exposes majority-class collapse; the class-weighted system recovers macro F1 by 301\% relative.}
\label{tab:task_b_results}
\centering
\footnotesize
\setlength{\tabcolsep}{3pt}
\begin{tabular}{llccc}
\toprule
\textbf{System} & \textbf{Sp.} & \textbf{Acc} & \textbf{Mac. F1} & \textbf{Wtd. F1} \\
\midrule
\multirow{2}{*}{Diagnostic base.}
 & Val  & 0.884 & 0.086 & 0.876 \\
 & Test & 0.884$^\dagger$ & 0.086$^\dagger$ & 0.876$^\dagger$ \\
\midrule
Class-weighted & Val & 0.787 & 0.345 & 0.836 \\
\bottomrule
\multicolumn{5}{l}{$^\dagger$Identical to val; see \S\ref{sec:results_b} for explanation.}
\end{tabular}
\end{table}

The stark contrast between accuracy (88.4\%) and macro F1 (0.086) in Table~\ref{tab:task_b_results} reveals systematic minority-class failure. The identical validation and test metrics are not coincidental: with both splits sharing the same 88.4\% human proportion, a near-constant majority predictor yields identical aggregate scores on each split, confirming that no genuine per-sample discrimination is occurring. A naive majority-class baseline achieves identical accuracy at macro F1 = 0.091, meaning the unweighted model \emph{performs below} the trivial baseline on the metric that matters.

\paragraph{Class-Weighted System.} Applying inverse-frequency class weighting and training for 3 epochs substantially improves macro F1 to 0.345 (+301\% relative) at the cost of majority-class accuracy (78.7\%). This confirms that imbalance remediation is both necessary and achievable. The accuracy decrease reflects the model abandoning its majority-class bias each point of accuracy lost corresponds to an LLM-generated sample now correctly attributed rather than collapsed into the human class. More aggressive strategies such as focal loss, contrastive pretraining, and class-balanced sampling are discussed in Section~7 as the most promising next steps.

\section{Analysis}

\subsection{Embedding Space Visualization}
\begin{figure}[H]
\centering
\includegraphics[width=\columnwidth]{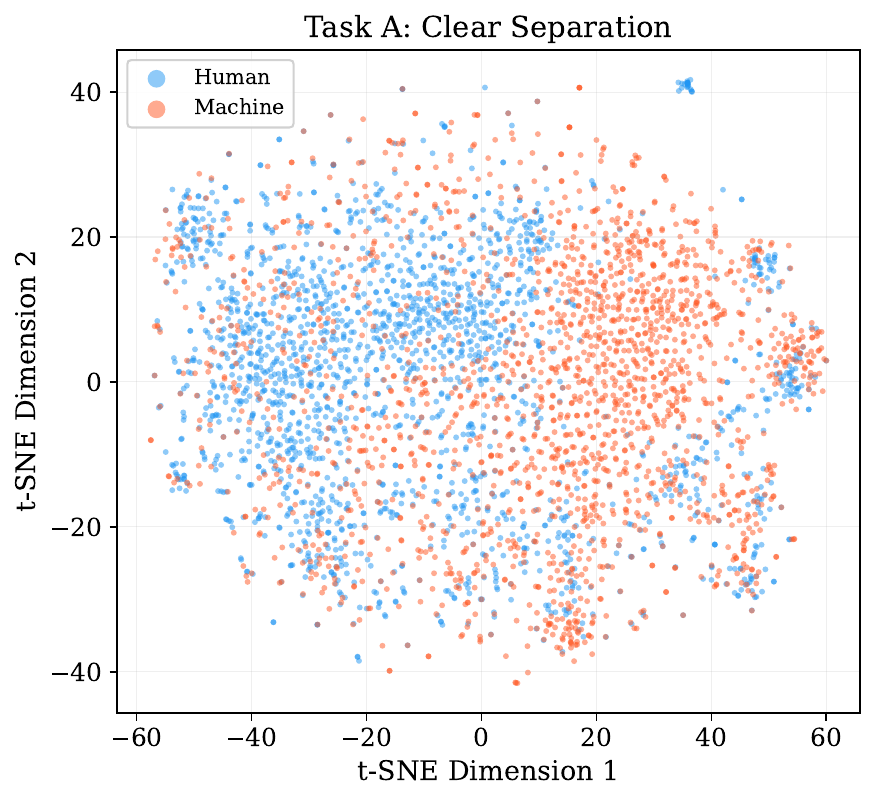}
\vspace{0.2cm}
\includegraphics[width=\columnwidth]{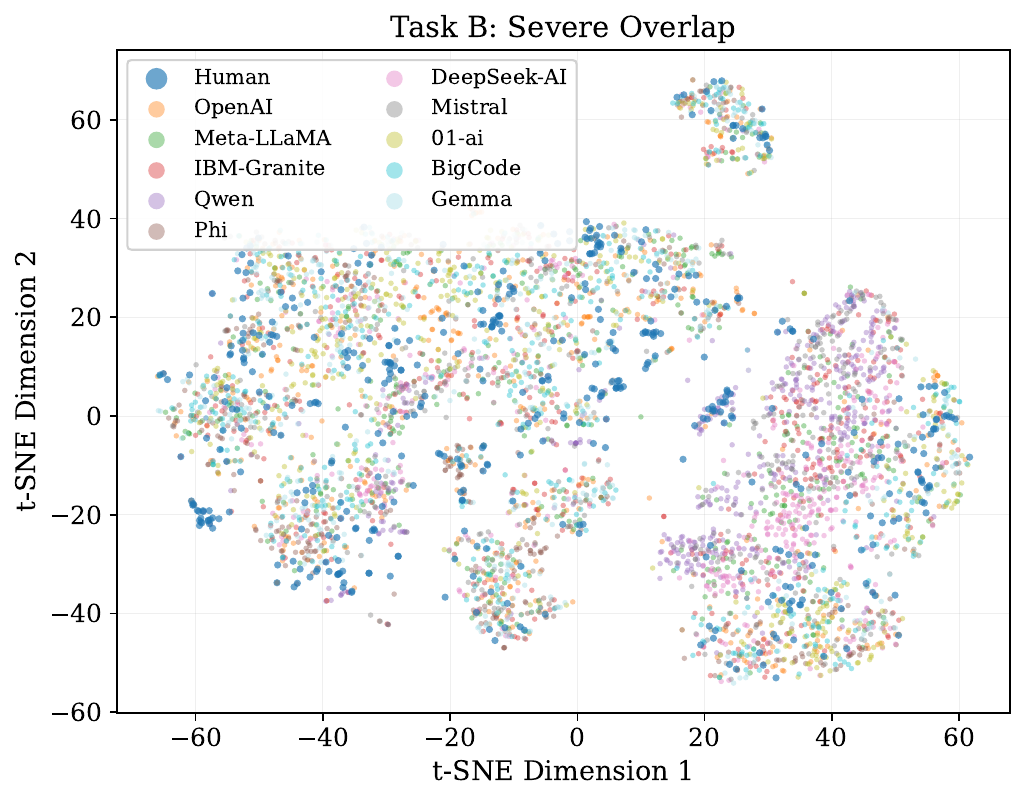}
\caption{Empirical t-SNE projections (perplexity=30, seed=42) of pretrained \texttt{[CLS]} embeddings extracted from held-out validation samples (2{,}000 per class for Task~A; up to 500 per class for Task~B). \textbf{Top}: pretrained UniXcoder embeddings for Task~A form separable human/machine clusters. \textbf{Bottom}: pretrained CodeBERT embeddings for Task~B show extensive class overlap, illustrating why fine-grained generator attribution collapses under standard fine-tuning.}
\label{fig:tsne}
\end{figure}

Figure~\ref{fig:tsne} provides empirical insight via t-SNE projections of pretrained \texttt{[CLS]} embeddings, UniXcoder for Task~A and CodeBERT for Task~B extracted from held-out validation samples.

\paragraph{Task A.} Human and machine code form distinct, well-separated clusters, consistent with the near-perfect validation classification.

\paragraph{Task B.} All classes overlap substantially. LLM families are not only indistinguishable from each other but also from human code in the embedding space, suggesting that CodeBERT's representations lack the granularity for fine-grained generator attribution.

\subsection{Why Does LLM-Generated Code Look Similar?}

We hypothesize three contributing factors. First, \textbf{shared training data}: most LLMs train on overlapping code corpora (GitHub, StackOverflow), learning comparable patterns. Second, \textbf{architectural homogeneity}: most generators use decoder-only transformer architectures with similar tokenization and generation strategies. Third, \textbf{problem constraints}: competitive programming solutions have limited valid approaches, reducing inter-generator stylistic variance.

\subsection{Ablation Study}
\label{sec:ablation}

\begin{table}[t]
\caption{Ablation study on validation sets. Sequence length significantly impacts Task~A; class weighting with full training yields the largest Task~B improvement (+301\% relative macro F1) at the cost of majority-class accuracy.}
\centering
\small
\begin{tabular}{lccc}
\toprule
\textbf{Configuration} & \textbf{Acc} & \textbf{M-F1} & \textbf{$\Delta$F1} \\
\midrule
\multicolumn{4}{l}{\textit{Subtask A -- Validation (Binary)}} \\
\midrule
Full system (512 tok) & 99.3 & 99.3 & --- \\
\quad $-$ 256 tokens & 99.0 & 99.0 & $-$0.3 \\
\quad $-$ 128 tokens & 98.1 & 98.1 & $-$1.2 \\
\quad $-$ 64 tokens & 94.7 & 94.6 & $-$4.7 \\
\quad $-$ No KL loss & 98.8 & 98.8 & $-$0.5 \\
\quad $-$ No warmup & 99.1 & 99.1 & $-$0.2 \\
\midrule
\multicolumn{4}{l}{\textit{Subtask B -- Validation (Multi-class)}} \\
\midrule
Full system (512 tok) & 88.4 & 8.6 & --- \\
\quad $-$ 256 tokens & 88.2 & 8.2 & $-$0.4 \\
\quad $-$ 128 tokens & 87.9 & 7.8 & $-$0.8 \\
\quad $-$ 1 epoch & 88.3 & 7.9 & $-$0.7 \\
\quad $+$ Class weights (3 ep.) & 78.7 & 34.5 & $+$25.9 \\
\bottomrule
\end{tabular}
\label{tab:ablation}
\end{table}

Table~\ref{tab:ablation} reveals several insights from validation-set experiments:

\paragraph{Sequence Length.} Reducing from 512 to 64 tokens drops Task~A macro F1 by 4.7 points, confirming that stylistic signals are distributed across moderate-length contexts and cannot be captured from short prefixes alone.

\paragraph{KL Consistency Loss.} Removing the symmetric KL regularization decreases validation F1 by 0.5 points, with a larger expected impact on out-of-distribution test performance since this component specifically targets cross-lingual robustness.

\paragraph{Class Weighting Trade-off.} Inverse-frequency class weighting trained for 3 epochs improves Task~B macro F1 from 8.6 to 34.5 (+301\% relative) but reduces accuracy from 88.4\% to 78.7\%. This confirms that imbalance remediation is both necessary and effective: the accuracy decrease reflects the model abandoning its majority-class bias in favour of genuine multi-class discrimination.

\subsection{Error Analysis}

\begin{figure}[t]
\small
\begin{tabular}{|p{0.95\columnwidth}|}
\hline
\textbf{Example 1: Human $\rightarrow$ Machine (False Positive)} \\
\texttt{def solve(n, arr):} \\
\texttt{~~~~return sorted(arr)[:n//2]} \\
\textit{Analysis}: Unusually concise human code misclassified as AI-generated due to lack of comments and one-liner style. \\
\hline
\textbf{Example 2: Machine $\rightarrow$ Human (False Negative)} \\
\texttt{\# This function calculates...} \\
\texttt{def calc\_sum(numbers):} \\
\texttt{~~~~total = 0  \# initialize} \\
\texttt{~~~~for num in numbers:} \\
\texttt{~~~~~~~~total += num} \\
\texttt{~~~~return total} \\
\textit{Analysis}: GPT-4 output with verbose inline comments mimics human tutorial style, evading detection. \\
\hline
\end{tabular}
\caption{Representative misclassifications. Both errors arise from style transfer: the model relies on surface-level stylistic patterns rather than deeper structural signals.}
\label{fig:error_examples}
\end{figure}

Figure~\ref{fig:error_examples} presents characteristic errors, both arising from \emph{style transfer}: concise human code resembles AI-typical efficiency, while verbose AI output mimics human pedagogical patterns. This confirms that the detector relies primarily on surface-level stylistic features, motivating the incorporation of deeper structural signals (ASTs, control flow) in future work.

\section{Discussion and Future Work}

Our results reveal two distinct regimes in machine-generated code detection. Binary human-vs-machine classification is highly effective: pretrained code models capture fundamental distributional differences, and our multi-view regularization framework achieves strong generalization even under severe language and domain shift. Fine-grained generator attribution, however, remains an open challenge. The embedding-space overlap across LLM families (Figure~\ref{fig:tsne}) combined with extreme class imbalance creates a setting where standard fine-tuning is insufficient.

Several directions appear promising for addressing attribution. Contrastive learning methods \citep{xu2025contrastive} that explicitly maximize inter-class separation may overcome embedding overlap. Incorporating structural features such as AST patterns and control flow graphs \citep{allamanis2018learning, guo-etal-2021-graphcodebert} could capture generator-specific signatures invisible at the token level. Focal loss \citep{lin2017focal} provides a more principled alternative to class weighting by dynamically downweighting easy majority-class examples; given that class weighting alone yields +301\% relative macro F1, focal loss would be expected to offer further improvement. Class-balanced sampling and few-shot meta-learning may additionally improve minority-class representations given the limited per-class training data.

\section*{Limitations}

We identify the following limitations of our work:

\begin{enumerate}
    \item \textbf{Model scope}: We evaluate only encoder-based architectures (UniXcoder, CodeBERT). Decoder-based or encoder--decoder models (e.g., CodeT5) may exhibit different generalization properties for both subtasks.

    \item \textbf{Incomplete Subtask~B exploration}: While our class-weighted system substantially improves macro F1 (+301\% relative), techniques such as focal loss, contrastive pretraining, and class-balanced sampling remain unexplored and would likely yield further gains.

    \item \textbf{Training data skew}: The Python-dominant training set (91\%) limits the strength of our cross-language generalization claims. Performance on truly low-resource languages (Go: 0.5\% of test) remains undertested.

    \item \textbf{Domain coverage}: Both training and a majority of evaluation samples originate from competitive programming. Generalization to production codebases, library code, or multi-file projects is unknown.

\section*{Ethics Statement}

This work targets high-stakes contexts such as 
academic integrity and hiring evaluation. our system's 0.760 human recall (Table~\ref{tab:task_a_results}) indicates 
that human review is essential before any 
consequential decision. False positives may 
disproportionately affect developers whose 
coding style diverges from the training 
distribution. We release our code to encourage 
reproducible, transparent, and responsible 
deployment.
\end{enumerate}

\clearpage
\bibliography{custom}

\end{document}